\documentclass[twocolumn,aps,showpacs,preprintnumbers,amsmath,amssymb]{revtex4}
\usepackage{graphicx}

\begin{document}

\title{ Towards Heisenberg Limit in Magnetometry with Parametric Down Converted Photons}

\author{Aziz Kolkiran}
\author{G. S. Agarwal}
\affiliation{Department of Physics, Oklahoma State University,
Stillwater, OK - 74078, USA}
\date{\today}

\begin{abstract}
Recent theoretical and experimental papers have shown how one can
achieve Heisenberg limited measurements by using entangled
photons. Here we show how the photons in non-collinear down
conversion process can be used for improving the sensitivity of
magneto-optical rotation by a factor of four which takes us
towards the Heisenberg limit. Our results apply to sources with
arbitrary pumping. We also present several generalizations of
earlier results for the collinear geometry. The sensitivity
depends on whether the two-photon or four-photon coincidence
detection is used.
\end{abstract}
\pacs{00000} \maketitle
\section{Introduction}
Parametric down conversion is a process that is used to produce
light possessing strong quantum features. Photon pairs generated
by this process show entanglement with respect to different
physical attributes such as time of arrival \cite{Friberg 1985}
and states of polarization \cite{Kwiat 1995}. They are
increasingly being utilized for very basic experiments to test the
foundation of quantum mechanics and to do quantum information
processing \cite{Mandel 1995, Zeilinger 1999,Kwiat 1995}. It is
also recognized that entangled photon pairs could be useful in
many practical applications in precision metrology involving e.g.
interferometry \cite{Caves 1981,Yurke 1986,Dowling 1998,Holland
1993}, imaging \cite{Abouraddy 2001,Pittman 1995}, lithography
\cite{Boto 2000,Agarwal 2001,Bjork 2001,Dangelo 2001} and
spectroscopy \cite{Agarwal 2003}. There is a proposal \cite{Lee
2002} to use electromagnetic fields in $NOON$ states to improve
the sensitivity of measurements by a factor of $N$. Some
implementations of this proposal exist \cite{Walther 2004}. In
particular, the use of photon pairs in interferometers allows
phases to be measured to the precision in the Heisenberg limit
where uncertainty scales as $1/N$ \cite{Giovannetti 2004} as
compared to the shot noise limit where it scales as $1/\sqrt{N}$.
This means that for large number of particles, a dramatic
improvement in measurement resolution should be possible.

In this paper we present an analysis of how parametric down
converted photons could be very useful in getting better
spectroscopic information about the medium. We demonstrate how the
improvement in magneto-optical rotation (MOR) of light could be
realized by employing two different schemes with collinear and
non-collinear down conversion geometry in compared to use of
coherent light. We calculate the resolution that can be achieved
in the MOR's both by use of coherent light and down converted
light. We discuss that the Heisenberg limit \cite{Ou 1997} could
be reached in magnetometry by the use down converted light.

\section{MOR using coherent light source} Consider a single mode
coherent light travelling in the z-direction and a linear isotropic
medium made anisotropic by the application of the magnetic field
$\mathbf{B}$ in the z-direction. The incident field can be written
in the form
\begin{equation}
{\mathbf E}(z,t)=\exp(-i\omega
t+ikz)(\hat{x}\varepsilon_x+\hat{y}\varepsilon_y)+c.c.
\end{equation}
The medium is described by the frequency and magnetic field
dependent susceptibilities $\chi_{\!\pm}(\omega)$. That means
horizontally and  vertically polarized components of the incident
light will rotate on travelling the medium of length $l$ and the
field at the exit can be written as
\begin{equation}
{\bf E}(l,t)=\exp(-i\omega
t+ikl)(\hat{x}\varepsilon_{xl}+\hat{y}\varepsilon_{yl})+c.c.
\end{equation}
The rotation of the horizontal and vertical components can be
expressed by the relations
\begin{equation}
\left(\begin{array}{c}\varepsilon_{xl}\\\varepsilon_{yl}\end{array}\right)=R
\left(\begin{array}{cc}\varepsilon_{x}\\\varepsilon_{y}\end{array}\right)
\end{equation}
where
\begin{eqnarray}\label{rotation}
R &=& e^{i\theta_{\!+}}
e^{i\frac{\theta}{2}}\left(\begin{array}{cc}\cos\frac{\theta}{2} &
-\sin\frac{\theta}{2} \\ \sin\frac{\theta}{2} &
\cos\frac{\theta}{2} \end{array}\right), \\
\theta &=& kl(\chi_{\!+} - \chi_{\!-}),\label{mor}\\
\theta_+\!\!\!&=& kl\chi_{\!+}.
\end{eqnarray}

\begin{figure}
 \scalebox{0.95}{\includegraphics{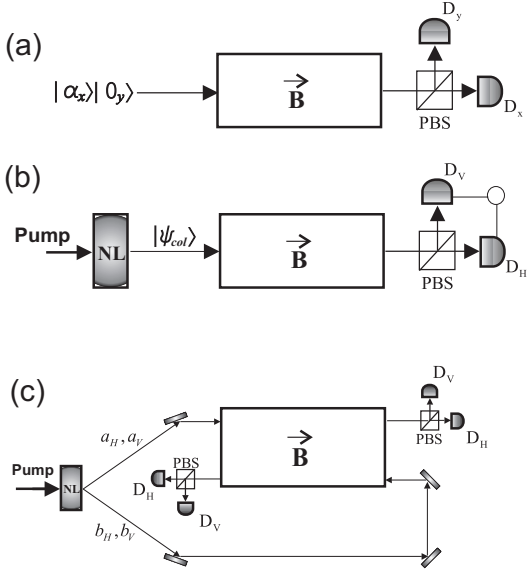}}
 \caption{\label{setup}The setup for the Magneto-optical rotation of light by using (a) coherent source, type-II PDC photons with (b) collinear and
 (c) non-collinear geometry.}
 \end{figure}

The corresponding quantum mechanical description can be obtained
 by replacing the classical amplitudes $\varepsilon_x$ and
$\varepsilon_y$ by the annihilation operators $a_x$ and $a_y$
respectively. For measurements with coherent sources one can look at
the intensities of the $x$ and $y$ components of the output when the
input is $x$ polarized with coherent state amplitude $\alpha_x$ (See
Fig. \ref{setup}(a). Then the measured quantities will be
\begin{eqnarray}
I_{xl}&=&\langle {a_{xl}}^{\dag}a_{xl}\rangle=|\alpha_x|^2
\cos^2\frac{\theta}{2},\label{coherent1}\\
I_{yl}&=&\langle {a_{yl}}^{\dag}a_{yl}\rangle=|\alpha_x|^2
\sin^2\frac{\theta}{2}.\label{coherent2}
\end{eqnarray}
One can estimate the minimum detectable rotation angle $\theta_m$
by looking at the fluctuations $\Delta N_d$ in the photon number
difference between horizontal and vertical photons, where the
number difference operator is given as
$N_d={a_{yl}}^{\dag}a_{yl}-{a_{xl}}^{\dag}a_{xl}$. This expression
is calculated to be $(\Delta N_d)^2=|\alpha_x|^2\sin^2\theta$ and
since the fluctuation noise is 1 we obtain $\theta_m\approx
1/\sqrt{\langle N\rangle}$ where $\langle N\rangle$ is the mean
number of input photons which is equal to $|\alpha_x|^2$.
\section{MOR using collinear type-II PDC and two-photon coincidence}
We now discuss how the results (\ref{coherent1}) and
(\ref{coherent2}) are modified if we work with down-converted
photons. We first consider the collinear case shown in Fig.
\ref{setup}(b). The state produced in collinear PDC can be written
by
\begin{equation}
|\psi_{col}\rangle=\frac{1}{\cosh
r}\sum_{n=0}^{\infty}(-e^{i\phi}\tanh r)^n |n\rangle_H
|n\rangle_V.
\end{equation}

The value of the parameter $r$ and the phase $\phi$ are related to
the pump amplitude of the nonlinear crystal that is used in the
down conversion process and the coupling constant between the
electromagnetic field and the crystal. Note that the state
$|\psi_{col}\rangle$ is a superposition of $n$ photon pairs of
horizontally and vertically polarized modes. Inside the medium,
these modes rotate with the same rotation matrix $R$ given in Eq.
(\ref{rotation}):
\begin{equation}
\left(\begin{array}{c}a_{Hl}\\a_{Vl}\end{array}\right)=R
\left(\begin{array}{cc}a_H\\a_V\end{array}\right).
\end{equation}
One can measure the intensity of each mode:
\begin{eqnarray}
I_H&\equiv&\langle {a_{Hl}}^{\dag}a_{Hl}\rangle=\sinh^2r\nonumber\\
&=&\langle {a_{Vl}}^{\dag}a_{Vl}\rangle\equiv I_V.
\end{eqnarray}
And the two-photon coincidence count is:
\begin{eqnarray}\label{co-PDC}
I_{HV}&\equiv &\langle
{a_{Hl}}^{\dag}{a_{Vl}}^{\dag}a_{Hl}a_{Vl}\rangle\nonumber\\
&=&\cos^2\theta\sinh^2r\cosh^2r+\sinh^4r
\end{eqnarray}

\begin{figure}
 \scalebox{0.70}{\includegraphics{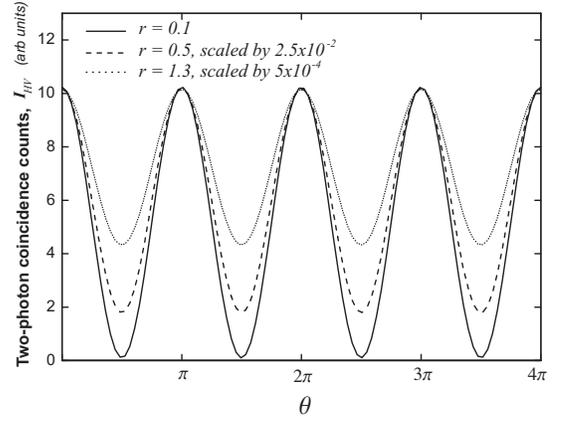}}
 \caption{\label{collinear-twophoton}The MOR plot of two-photon coincidence counts defined by the Eq. (\ref{co-PDC}) in collinear type-II PDC.
 $r$ is the interaction parameter that defines the pumping strength
 used in the production of down converted photons and $\theta=kl(\chi_+-\chi_-)$. }
 \end{figure}

\begin{figure}
 \scalebox{0.70}{\includegraphics{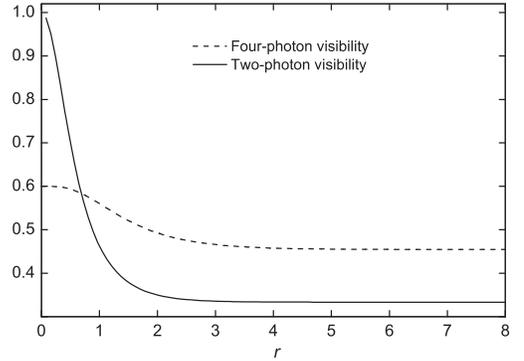}}
 \caption{\label{visibility}The visibility of two-photon and four-photon counts defined by the Eqs.
 (\ref{co-PDC}) and (\ref{4photon in detector}) respectively.}
 \end{figure}

Note the difference between Eqs. (\ref{coherent1}) and
(\ref{co-PDC}). With collinearly down-converted photons we measure
a rotation angle that is twice as large compared with the angle
for a coherent input. For $r\ll 1$ we obtain the same result as
given in \cite{Agarwal 2003}. The fringe pattern and the
visibility is given in Figs. \ref{collinear-twophoton} and
\ref{visibility}. One can calculate the minimum detectable
rotation angle again by looking at the fluctuations in the photon
number difference $N_d$. This is given by $(\Delta
N_d)^2=4\sinh^2r\cosh^2\sin^2\theta=(1+\langle N\rangle)\langle
N\rangle\sin^2\theta\approx{\langle N\rangle}^2\sin^2\theta$ for
large $\langle N\rangle$ where $\langle N\rangle=2\sinh^2r$.
Making $(\Delta N_d)\sim 1$ \cite{Ou 1997} we get $\theta_m\approx
1/\langle N\rangle$. Note that the sensitivity of this quantity is
also improved by a factor of $1/\sqrt{N}$.
\section{MOR using non-collinear type-II PDC and four-photon coincidence} Next, we discuss the
non-collinear PDC case. We have found an arrangement shown in Fig.
\ref{setup}(c) which is especially attractive for improving
sensitivity. The entangled photons are coming in two different
spatial modes, $a$ and $b$. While one mode (say $a$) is going
parallel to $\mathbf{B}$ inside the medium, the other is going
anti-parallel to it. At the exit we separate the $H$ and $V$ modes
by polarizing beam splitters. The state of the input photons can
be written in the form \cite{Kok 2000}
\begin{equation}
|\psi_{non}\rangle=\frac{1}{\cosh^2r}\sum_{n=0}^{\infty}\sqrt{n+1}(\tanh
r)^n|\psi_n\rangle,
\end{equation}
where
\begin{equation}\label{state2b}
|\psi_n\rangle=\frac{1}{\sqrt{n+1}} \sum_{m=0}^n (-1)^m
|n-m\rangle_{a_H}|m\rangle_{a_V}|m\rangle_{b_H}|n-m\rangle_{b_V}.
\end{equation}
Here $|m\rangle_{a_V}$ represents $m$ vertically polarized photons
in mode $a$. Inside the medium, $``+"$ and $``-"$ polarization
components of the modes $a$ and $b$ gain phases $kl\chi_{\!+}$ and
$kl\chi_{\!-}$ respectively. Thus we can write an effective
Hamiltonian for the evolution of the state $|\psi_{non}\rangle$
inside the medium as follows:
\begin{equation}\label{medium}
H_{med}=\chi_{\!+}{a_{+}}^{\dag}a_{+}+
\chi_{\!-}{a_{-}}^{\dag}a_{-}- \chi_{\!+}{b_{+}}^{\dag}b_{+}-
\chi_{\!-}{b_{-}}^{\dag}b_{-},
\end{equation}
where
\begin{equation}
a_{\pm}=\frac{1}{\sqrt{2}}(a_H\pm ia_V),\quad
b_{\pm}=\frac{1}{\sqrt{2}}(b_H\pm ib_V).
\end{equation}
\begin{figure}
 \scalebox{0.50}[0.50]{\includegraphics{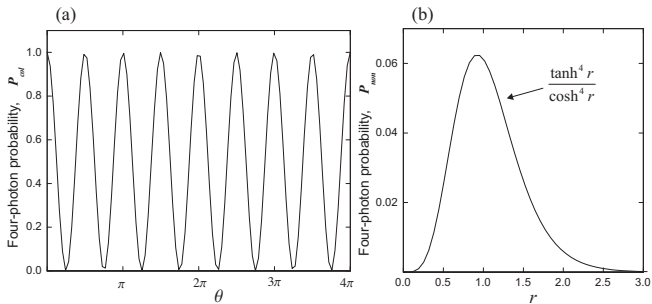}}
 \caption{\label{noncollinear-4photon}(a) The normalized four-photon probability defined in Eq. (\ref{4photon1}) , and (b) its envelope {\it
 wrt} the interaction parameter $r$ in the non-collinear geometry.}
 \end{figure}
\begin{figure}
 \scalebox{0.50}[0.50]{\includegraphics{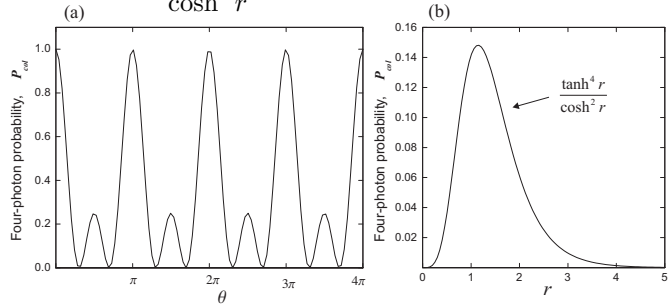}}
\caption{\label{collinear-4photon_a}(a) The normalized four-photon
probability defined in Eq. (\ref{4photon2}), and (b) its envelope
{\it wrt} the interaction parameter $r$ at the exit ports of PBS
in the collinear geometry.}
 \end{figure}
The minus sign in front of the $b_{\pm}$ modes comes from the fact
that they are travelling anti-parallel to the $\mathbf B$ field
inside the medium. Then one can calculate the probability of
detecting four photons in each mode as:
\begin{eqnarray}\label{4photon1}
P_{non}&=&|\langle
1_{a_H}1_{a_V}1_{b_H}1_{b_V}|\exp(-itH_{med})|\psi_{non}\rangle|^2\nonumber\\
&=&\frac{\tanh^4r}{\cosh^4r}\cos^2(2\theta),
\end{eqnarray}
where $t$ is the duration for the state to evolve inside the
medium. Note that this four-photon probability has the rotation
angle that is four-times as large compared with the angle for a
coherent input. The fringe pattern with respect to $\theta$ and
the probability distribution with respect to $r$ are shown in Fig.
\ref{noncollinear-4photon} (a) and (b).

Next we also examine the four-photon  probability in the collinear
case. The probability of finding two $H$-photons and two
$V$-photons at the exit ports of the polarizing beam splitter is
given by:
\begin{eqnarray}\label{4photon2}
P_{col}&=&|\langle2_{a_H}2_{a_V}|\exp(-itH_{med})|\psi_{col}\rangle|^2\nonumber\\
&=&\frac{\tanh^4r}{\cosh^2r}\frac{1}{16}[1+3\cos(2\theta)]^2,
\end{eqnarray}
where we take $H_{med}=\chi_{\!+}{a_{+}}^{\dag}a_{+}+
\chi_{\!-}{a_{-}}^{\dag}a_{-}$ because of the collinear geometry.
The normalized plot of this quantity with respect to the
magneto-optical rotation angle $\theta$ and the envelope of the
probability with respect to $r$ are shown in Fig.
\ref{collinear-4photon_a} (a) and (b).

\begin{figure}
 \scalebox{0.70}{\includegraphics{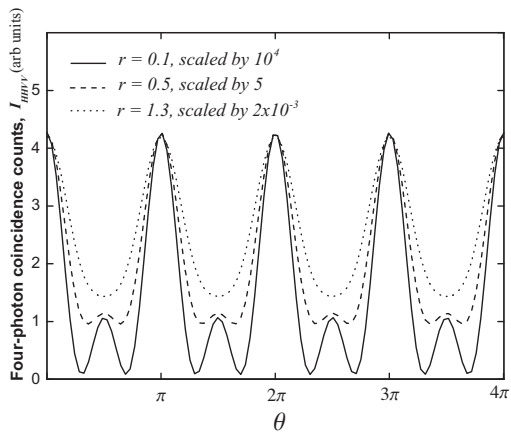}}
 \caption{\label{collinear-4photon}Four-photon coincidence counts defined in Eq. (\ref{4photon in detector})
 with different interaction parameter values in the collinear geometry.}
 \end{figure}

On the other hand one can also calculate the coincidence counts of
four photons two-by-two at each detector as given by Glauber's
higher order correlation functions:
\begin{widetext}
\begin{eqnarray}
I_{HHVV}&=&\langle
{a_{Hl}}^{\dag 2}{a_{Vl}}^{\dag 2}{a_{Hl}}^2{a_{Vl}}^2\rangle\nonumber\\
&=&(3\cos^2\theta-1)^2\sinh^4r\cosh^4r
+4(3\cos^2\theta+1)\sinh^6r\cosh^2r+4\sinh^8r.\label{4photon in
detector}
\end{eqnarray}
\end{widetext}
The plot of this quantity for different values of the interaction
parameter $r$ and the visibility are shown in Figs.
\ref{collinear-4photon} and \ref{visibility}.
Note the distinction between Eqs. (\ref{4photon2}) and
(\ref{4photon in detector}) which is a reflection of what the
detector is set to measure as we explain now. The former is the
probability of the state $|\psi_{col}(t)\rangle$ to be projected
onto the particular four-photon subspace $|22\rangle$ i.e.
$Tr[|22\rangle\langle22|\rho_{col}(t)]$ where
$\rho_{col}(t)=U|\psi_{col}\rangle\langle\psi_{col}|U^{\dag}$ and
$U$ is the unitary operator that represents the evolution of the
state by the Hamiltonian $H_{med}$ in the collinear geometry. On
the other hand, coincidence counting of four-photons at the
detectors $D_H$ and $D_V$ (see Fig. \ref{setup}(b)) is represented
by the expectation value
 $\langle{a_{H}}^{\dag 2}{a_{V}}^{\dag 2}{a_{H}}^2{a_{V}}^2\rangle=
 Tr[{a_{H}}^{\dag 2}{a_{V}}^{\dag
 2}{a_{H}}^2{a_{V}}^2\rho_{col}(t)]$. Note here that the operator  ${a_{H}}^{\dag 2}{a_{V}}^{\dag
 2}{a_{H}}^2{a_{V}}^2$ has the spectral decomposition $\sum_{nm}^{\infty}C_{nm}|nm\rangle\langle
 nm|$ and obviously it contains the projectors of all $(n+m)$-photon
 subspaces with nonzero coefficients $C_{nm}$. Therefore four-photon
 counting process at detectors includes not only $|22\rangle$ but all other states $|nm\rangle$ in $|\psi_{col}(t)\rangle$.
 Here the state $|nm\rangle$ represents $n$ and $m$ photons in the
 $a_H$ and $a_V$ modes respectively.


\section{Conclusion} We showed that the use of
non-collinear type-II PDC light in MOR's increases the sensitivity
by a factor of four in comparison to coherent light. We also give
an argument that minimum rotation uncertainty scales to Heisenberg
limit by the use of down converted photons. It should be noted
that Heisenberg limit should be understood as an approximate limit
at a large mean photon number, that is, the rotation uncertainty
approaches the order of $1/\langle N\rangle$ for large $\langle
N\rangle$ \cite{Holland 1993}. The regime with an interaction
parameter value of $r=1.3$ has already been reached in the
experiment \cite{Eisenberg 2004} giving entanglement of $12$
photons and an evidence also was given for entanglement up to 100
photons.
\appendix
\section{Four-photon probability} In this appendix we show the
details of the calculation leading to the result given in Eq.
(\ref{4photon1}). One can obtain the result first by solving the
Schr\"odinger equation for the state
$|1_{a_H}1_{a_V}1_{b_H}1_{b_V}\rangle$ in the four-photon subspace
of the electromagnetic field and having the inner product with the
state $|\psi\rangle_{non}$. Since the parts of the Hamiltonian
 having $a$ and $b$ modes commute, we can solve the
Schr\"odinger equation for the states $|1_{a_H}1_{a_V}\rangle$ and
$|1_{b_H}1_{b_V}\rangle$ separately. Let us start with a general
time-dependent state in the  $a_H$ and $a_V$ modes which contain two
photons totally;
\begin{equation}
|\phi(t)\rangle=c(t)|20\rangle+d(t)|02\rangle+f(t)|11\rangle
\end{equation}
with the initial condition $|\phi(0)\rangle=|11\rangle$. Solving
the Schr\"odinger equation by using the effective Hamiltonian
$H=\chi_{\!+}{a_{+}}^{\dag}a_{+}+ \chi_{\!-}{a_{-}}^{\dag}a_{-}$
gives us the result
\begin{eqnarray}
|\phi(t)\rangle&=&e^{-it\chi}[\frac{1}{\sqrt{2}}\sin(\Omega
t)|20\rangle-\frac{1}{\sqrt{2}}\sin(\Omega t)|02\rangle \nonumber \\
&+& \cos(\Omega t)|11\rangle]
\end{eqnarray}
where $\chi=\chi_{\!+} + \chi_{\!-}$ and $\Omega=\chi_{\!+} -
\chi_{\!-}$. For a medium of length $l$, the angle $\Omega t$
corresponds to the MOR angle $\theta$ which is given in Eq.
(\ref{mor}). The solution for the state $|1_{b_H}1_{b_V}\rangle$ can
be obtained just by replacing $\theta$ by $-\theta$ because the
direction of propagation of the $b$ modes are opposite to that of
$a$ modes inside the medium. This is the reason that the part of the
effective Hamiltonian for the $b_{\pm}$ modes takes minus sign in
Eq. (\ref{medium}). Consequently we can write the solution of the
Schr\"odinger equation for the state
$|1_{a_H}1_{a_V}1_{b_H}1_{b_V}\rangle$ as:
\begin{widetext}
\begin{eqnarray}
\exp(-itH_{medium})|1_{a_H}1_{a_V}1_{b_H}1_{b_V}\rangle &=&
\exp(-it\chi)\left[\frac{1}{\sqrt{2}}\sin\theta|20\rangle-
\frac{1}{\sqrt{2}}\sin\theta|02\rangle+\cos\theta|11\rangle
\right]\nonumber\\
&\otimes&
\exp(-it\chi)\left[-\frac{1}{\sqrt{2}}\sin\theta|20\rangle+
\frac{1}{\sqrt{2}}\sin\theta|02\rangle+\cos\theta|11\rangle \right].
\end{eqnarray}
\end{widetext}
Taking the inner product of this with the state
$|\psi\rangle_{non}$ and having the absolute square gives us the
result given in Eq. (\ref{4photon1}).

The result given in Eq. ({\ref{4photon2}}) can be obtained by
following the same method given above.



 \end{document}